\documentclass{ws-rv961x669}

\usepackage{ws-rv-van}     
\usepackage{ws-rv-thm}     
\usepackage{subfigure}     

\newcommand{\be}{\begin{equation}}             
\newcommand{\ee}{\end{equation}}               
\newcommand{\ba}{\begin{eqnarray}}             
\newcommand{\ea}{\end{eqnarray}}               
\newcommand{\n}[1]{\label{#1}}
\newcommand{\hh}{,\hspace{0.5cm}}
\newcommand{\hhh}{,\hspace{0.2cm}}

\makeindex

\begin{document}

\chapter[Remarks on black hole entropy]{Remarks on black hole entropy\label{ra_ch1}}

\author{Valeri P. Frolov}

\address{Theoretical Physics Institute, University of Alberta, Edmonton,
Alberta, Canada T6G 2E1 \\
vfrolov@ualberta.ca\footnote{Affiliation footnote.}}

\begin{abstract}
The notion of black-hole entropy was introduced by Jacob Bekenstein in 1972. During past 45 years this subject was in the center of interests of the modern theoretical physics. In this paper we briefly discuss "puzzles" of the black-hole physics, connected with their entropy. We also demonstrate that when the standard energy conditions are violated entropy associated with the event horizon can have quite unexpected behavior,
\end{abstract}


\body


\section{Instead of Introduction}

Jacob Bekenstein became famous after he published in 1972 three papers on  black hole entropy and no-hair theorem \cite{Bekenstein:1972tm,Bekenstein:1971hc,Bekenstein:1973ur}. He was only 25 years old at that time. His work was very close to my scientific interests and I followed all his publications. But for the first time I met Jacob only in the middle of 90th at the conference in Moscow. By that time he was already a world-wide recognized expert in black holes. What surprised me at our first meeting, was that this famous scientist  turned out to be such a shy person, very friendly and widely open for discussions.

After this I have met Jacob at many conferences and had a lot of discussions with him. I remember that at one of the  Quantum gravity meeting at Utrecht we were sitting together with Jacob. After my talk he mentioned that it was interesting but I spoke too fast. His talk was after mine. At the end of it because of the shortage of time he was speaking even faster than me. We laughed together when I mentioned this to him.

On many occasions  I tried to convince Jacob, that black hole quantization, which he and Slava Muchanov have advocated, not necessary means that the spectrum of black hole radiation should be discrete. My argument was that in their model the discrete levels have  huge degeneracy and the interaction between these internal degrees of freedom can make the spectrum of radiation practically continuous. Jacob did not agree with me and proposed arguments, based on the gedanken experiments, supporting his point of view. Jacob could be a bit stubborn.

In 2016 there was a celebration of a hundred anniversary of black holes. We with Don Page were organizing a  meeting "Black Holes' New Horizons" devoted to this event, which was held in May 2016 in Oaxaca (Mexico). We wanted very much to see Jacob at our meeting and sent him in the beginning of March of 2015 our invitation letter. A few days later (on March 6) I got the following answer from Jacob:

{\it "Dear Valeri,

So kind of you to invite me to the workshop in Oaxaca. I would have liked to go, but I am, of late, skipping meetings which involve very long travel. And this would be a classic example of such. I imagine I will be missing a lot of fun; such is life. But I wish you and Don a successful organizing of what must be a very exotic meeting.  I hope you enjoy it. My very best wishes,

Jacob"}

Jacob Bekenstein passed away on August 16, 2015. To honor his memory we dedicated to him a special session at our Black Hole meeting. Bill Unruh and Barack Kol, who knew Jacob very well, spoke about their memories dedicated to Jacob's life.

\section{Black-hole entropy puzzles}

\subsection{Bekenstein-Hawking entropy of astrophysical black holes is huge}

In his groundbreaking papers \cite{Bekenstein:1972tm,Bekenstein:1973ur} Jacob Bekenstein claimed that a black hole should have entropy proportional to its surface area. He arrived to this conclusion by considering gedanken experiments with black holes. His conclusion was that this entropy  is $S_H=\beta A/l_{Pl}^2$, where the dimensionless coefficient $\beta$ is of order of 1. Hawking calculations of the quantum radiation of black holes \cite{Hawking:1974rv,hawking1975}
demonstrated that it is thermal and the temperature of a black hole of mass $M$ is
\be
T_{BH}={\hbar c^3\over 8\pi G M k_B}\, ,
\ee
where $k_B$ is the Boltzmann constant. This result allowed to fix the parameter $\beta$ in the Bekenstein's expression for the entropy, $\beta=1/4$. The exact expression for what is called now Bekenstein-Hawking entropy is\footnote{The standard subscript $BH$ used in the expression has double meaning. It is used as an abbreviation for the "Black Hole". It   can also be understood as the abbreviation for "Bekenstein-Hawking".}
\be\n{BHBH}
S_{BH}={A\over 4l_{Pl}^2}\, .
\ee

Already in his first papers, Bekenstein estimated the black hole entropy for stellar mass black holes, and found that is huge. In \cite{Bekenstein:1973ur} he wrote:

\begin{figure}[h]
\centerline{\includegraphics[width=2.5in]{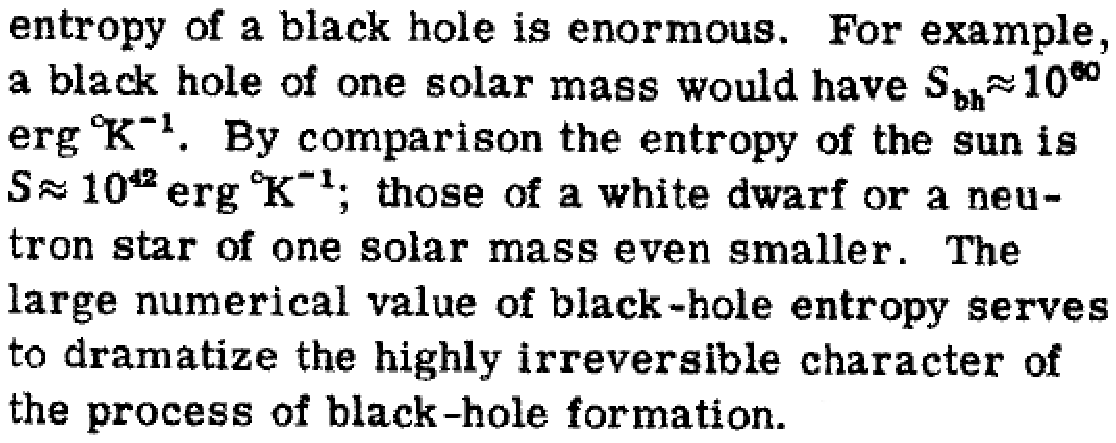}}
\label{BEK}
\end{figure}

Now we know that besides stellar mass black holes there also exist supermassive black holes with the mass $M\sim 10^6-10^{10} M_{\odot}$. The corresponding dimensionless entropy of these objects is $S_{BH}\sim 10^{89}-10^{97}$.

The papers \cite{BH_UNIV,UNIV_ENT} contain estimations of the contribution of matter and black holes to the entropy of the Universe. The table below taken from the paper \cite{BH_UNIV} summarizes its  results.
\begin{table}[htb]
\begin{center}
\begin{tabular}{c|c|c}
\hline
{\rm objects} & {\rm entropy} &  {\rm energy} \\
\hline
$10^{22}$ stars  &  $10^{79}$   &   $\Omega_{\rm stars} \sim 10^{-3}$  \\
relic neutrinos  &  $10^{88}$   &   $\Omega_{\rm \nu} \sim 10^{-5}$  \\
stellar heated dust  & $10^{86}$ & $\Omega_{\rm dust} \sim 10^{-3}$ \\
CMB photons  &  $10^{88}$   &   $\Omega_{\rm CMB} \sim 10^{-5}$  \\
relic gravitons  &  $10^{86}$   &   $\Omega_{\rm grav} \sim 10^{-6}$  \\
stellar BHs & $10^{97}$&$\Omega_{\rm SBH} \sim 10^{-5}$\\
single supermassive BH & $10^{91}$ & $10^7 M_\odot$ \\
$10^{11} \times 10^7 M_\odot $ SMBH & $10^{102}$ &  $\Omega_{\rm SMBH} \sim 10^{-5}$ \\
holographic upper bound & $10^{123}$ &  $ \Omega = 1$ \\
\hline
\end{tabular}
\end{center}
\caption{Entropies and energies . The table is taken from \cite{BH_UNIV} }
\label{tbl:I}
\end{table}

The main conclusion is that up to best of our present knowledge most of the entropy of our Universe is  in the form of the black-hole entropy. The following example also illustrates a peculiar property of black holes as thermodynamical systems. Let us assume that one kilogram of matter falls down into the black hole of mass $10^{10} M_{\odot}$. The change of the entropy in this process is  as large as $\Delta S_{BH}\sim 10^{77}$ and it is comparable with the thermal entropy of a single star.

The concept of black-hole entropy, introduced by Jacob Bekenstein, is now widely used. Google search shows that more than a half of million documents uses this notion. There are several puzzles of the black hole physics which are connected with this notion. Below we briefly discuss some of them.

\begin{figure}[ht]
\centerline{\includegraphics[width=1.7in]{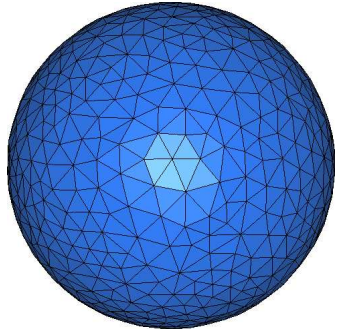}}
\caption{Sphere triangulation} \label{sphere}
\end{figure}

One can illustrate the key problem connected with the notion of the black-hole entropy as follows. Suppose one has a sphere of radius $R$. Its surface area is $A=4\pi R^2$. Let us consider triangulation of the sphere and assume the area of a single triangle is $s=l^2$ (see Figure~\ref{sphere}). The total number of the triangles is $N=A/s$. One can use such a sphere to encode information, for example, by coloring triangles by black and white colors. Total number of possible combinations is $2^N$, and the amount of information that one can transfer by using such a code is $\sim A/l^2$. The entropy corresponding to the loss of this information is $S\sim A/l^2$. This simple observation shows that in order to obtain $S_{BH}$ the size of elementary cells must be of order of the Planckian length. In other words, a plausible explanation of the statistical-mechanical origin of the black-hole entropy should appeal to quantum gravity.

\subsection{Black-hole entropy and entanglement}

In the interaction with surrounding matter a black hole behaves as a heated body with temperature $T_{BH}$ and entropy $S_{BH}$. Black holes obey laws, similar to the laws of the standard thermodynamics \cite{Bardeen:1973gs} (for a comprehensive discussion see e.g. reviews \cite{Wald:1999vt,Wald:2002mon}). A natural question which attracted a lot of attention is: Is the analogy of black holes with thermodynamical systems complete? Do there exist internal degrees of freedom responsible for the black-hole entropy? One of the first attempts to answer this question was proposed by Bombelli,  Koul, Lee and Sorkin (BKLS) \cite{Bombelli:1986rw}. They propose  to relate the black-hole entropy to the entanglement entropy of zero point fluctuations. A similar idea was proposed by t'Hooft \cite{tHooft:1990fkf} in his famous brick wall model. The BKLS idea was rediscovered in\cite{Srednicki:1993im,PhysRevD.48.4545} , where  the entanglement entropy of a free quantum field in the spacetime of a black hole was calculated. There were many interesting and important publications on this subject (see e.g. \cite{Susskind:1993if, Callan:1994py, Holzhey:1994we, Kabat:1994vj}). A nice review, which contains a detailed analysis of these and other publications and describes the "state of art" of the black hole entropy problem by 1994 can be found in the paper by Jacob Bekenstein \cite{Bekenstein:1994bc}.

The origin of such entanglement entropy can be understood as follows. The interior of the black hole is bounded by the  event horizon. For an external observer a regular at the horizon (vacuum) state of a quantum field can be described in terms of Rindler particles (RP). Each of these RP has a partner particle (PP) located inside the horizon and which is strongly correlated with it. A superposition of RP and PP is  a pure quantum state. Averaging over unobservable  by an external observer PP states results in the density matrix $\hat{\rho}_{RP}$ for RP states. It describes the system entanglement. The entanglement entropy is defined as $S=-Tr_{RP}(\hat{\rho}_{RP} \ln \hat{\rho}_{RP})$.

Some of the Rindler particles penetrate the potential barrier, surrounding the black hole. They form the Hawking radiation. However, the vast majority of Rindler particles are reflected by the potential barrier and  fall down inside the black hole. At the same time new RPs are injected by the black hole. As the result, the distribution of the RP outside of the horizon is stationary. The RPs during their (short) lifetime outside the black hole contribute to the entanglement entropy. The corresponding density matrix $\hat{\rho}_{RP}$ is thermal. This property is closely related with the thermality of the Hawking radiation. Since the vast majority of the Rindler quanta that contribute to the entropy are located in the very close vicinity of the horizon, this entropy is proportional to the surface area of the black hole.

The problem of this and similar approaches to the black-hole entropy is that this quantity is divergent. The divergence is a consequence of the assumption that the black hole surface is infinitely sharp. Certainly, this is an idealization. In fact, as a result of emission of the Hawking quanta, the horizon fluctuates. This effect of the horizon fluctuation was described in the remarkable paper by Jacob Bekenstein \cite{Bekenstein:1984book}.  In terms of proper distance, the scale of such fluctuations is of the order of the Planck length, $l_{Pl}$. By using this cut-off one obtains the finite expression for the black-hole entanglement entropy
\be\n{SBS}
S_{ent}=\beta S_{BH}\, ,
\ee
where the dimensionless parameter $\beta$ is of order of 1 \cite{PhysRevD.48.4545}.

There were several, more formal proposals, for making $S_{ent}$ finite. One of the most popular attempts was some analogue of the renormalization procedure. It was proposed to prescribe to a black hole (infinite) negative, so called "geometric",  entropy, in such a way that after adding it to the (infinite) positive entanglement entropy one obtains a finite answer, which is identified with $S_{BH}$. The weak point of this approach is that it does not allow one to explain black hole entropy by statistical-mechanical counting of some  black hole degrees of freedom. It introduces the notion of "geometric  entropy", which does not have any statistic-mechanical meaning.

\subsection{Universality problem}

The above described approach, in which the entropy of the black hole is related to existing physical fields,  has a  fundamental problem. Each of the fields gives independent (additive) contribution to $S_{ent}$. Hence, even if the proper physical mechanism of the cut-off is found, the factor $\beta$, which enters (\ref{SBS}),  depends on the number and characteristics of the fields that contribute to $S_{ent}$. In particular, it depends on the spectrum of  mass and  properties of  fundamental elementary particles, even of those that are not yet discovered.

This problem is closely related to another, even more fundamental problem: The universality of the black hole entropy $S_{BH}$. Consider a black hole with mass much larger than the Planckian mass. Such an object is classical and one can neglect  possible quantum corrections. In other words, in order to find the Bekenstein-Hawking entropy of the black hole it is sufficient to use the corresponding solution of the classical Einstein equations. At the same time, one is going to obtain the same answer by counting some microscopic degrees of freedom of some version of quantum gravity. In order to obtain the required value of $S_{BH}$ the corresponding constituents must be heavy, with mass comparable with the Planckian one. The calculations, based on the fundamental theory should correctly reproduce the result of the low energy classical gravity. In other words, there must exist a mechanism, that guarantees that the calculations based on counting of  states in the fundamental theory always  give the correct answer, $S_{BH}$, which is known from the low energy gravitational theory. This is a so called universality problem \cite{Frolov:1995pt,CARLIP_UNIVER,Carlip:2008rk,STROM,SAIDA,Carlip:2008rk}.

There are several approaches to counting the black-hole degrees of freedom and calculation of its entropy based on the some "fundamental" theory, such as string theory, loop gravity, induced gravity, etc. A  discussion and comparison of these approaches can be found in comprehensive reviews \cite{Carlip:2008wv,Carlip:2017uzi}. Common feature of these approaches is that the low energy gravity is an emergent phenomenon generated by fundamental quantum heavy  constituents (strings, loops, etc). In such a case the black-hole entropy is obtained by counting the number of states of such constituents.

\subsection{Information loss paradox}

Another black-hole puzzle is a so called information loss paradox. It can be formulated as follows. Suppose a black hole was formed as a result of the collapse of matter, which originally was in pure quantum  state. For example, it was a coherent state of  collapsing photons. After its formation black hole emits Hawking radiation and its mass decreases. If as result of this process the black hole disappears, the final state of the system would be radiation emitted by the black hole. In other words,  a pure state is transformed into a state described by the density matrix with non-zero entropy. This implies the violation of unitarity of quantum mechanics \cite{PhysRevD.72.084013}. Certainly, there is an option that the evaporation is not complete and some remnant would remain. If the initial mass of the black hole was large, this remnant must contain partners of all the particles emitted outside during Hawking evaporation. Since the initial state was pure, the entropy of these partners is the same as the total entropy of the emitted outside particles. In the classical description, this entropy is finally absorbed by the singularity inside the black hole. If in an UV complete version of Einstein gravity one or more new universes are created inside the black hole \cite{Frolov:1988vj,Frolov:1989pf}, they "inherit" the corresponding entropy.

The information loss paradox arises if for some reason this "scenario" is considered as unsatisfactory, and one insists that the evaporation of the black hole is complete and  no remnant is left. This point of view is adopted in the string theory. There exist large number of publications devoted to the unitarity problem in the black hole evaporation (see e.g. reviews \cite{Preskill:1992tc,Page:1994pt,Banks:1994ph,Thorlacius:1994ip,Strominger:1994tn,0264-9381-26-22-224001,Frolov:2014jva,Unruh:2017uaw}). For example, one can assume that (1) the information "comes out" with the Hawking radiation \cite{Page:1994pt}, or (2) the information is "stored"  in soft-modes near the horizon \cite{Hawking:2016msc,Hawking:2016sgy},
or (3) the information "comes out" at the final state of the evaporation (see e.g. \cite{Hawking:2014tga,Frolov:2014jva,Bardeen:2014uaa} and references therein).

There is another interesting question which is directly connected with the information loss paradox. Let us consider two black holes, which at  some moment of time $T$ have same mass $M$. Let us assume that they were formed as a collapse of matter in the pure state.  Let us also assume that one of these black hole was just formed, while the other one was formed long time ago, at time $T_0<T$. Its initial mass was $M_0>M$, and it lost the mass $\Delta M=M_0-M$ as a result of the Hawking evaporation. Before the mass of the second black hole becomes $M$ it emits radiation which has entropy $\sim \Delta S\sim M^2_0-M^2$. For example, if $M_0=\sqrt{2}M$, the emitted entropy $\Delta S$ is of the order of $S_{BH}(M)$. The time during which the surface area of the event horizon of an evaporating black hole is halved is often called  "Page time" \cite{PAGE_TIME}.

Evaporation of particle in the black hole exterior is accompanied by creation of its partner inside the black hole. By time $T$ the entropy of such partners is the same, as the entropy of the emitted particles, $\Delta S$. Thus the "internal states" of these two black hole, "young" and "old", with the same mass $M$ at time $T$ is different. If this information returns to the external observer during the last stage of their complete evaporation, this phase duration should be different for two black holes. A result reported  by Gregory Vilkovisky   some time ago \cite{Vilkovisky:2005cz,Vilkovisky:2005db,Vilkovisky:2008vv}   might be interesting in connection with this problem.

The number of papers on this subject is fast growing. This means that at the moment we do not have a solution of this puzzle, at least the one which satisfies most the gravity-string community.

\subsection{Where black-hole entropy is located?}

In the discussion of black holes many different definitions of their horizon were proposed, such as: event horizon, apparent horizon, isolated horizon etc. Most of the statistical-mechanical calculations of the black hole entropy was done under an assumption that the black hole is either static or stationary. In this case different definitions of black hole horizon agree. And it is natural to associate the black hole entropy with the event horizon. However, if the black hole is not stationary and its mass and shape depend on time the  question where black hole entropy is located becomes non-trivial. Difference between different definitions of the horizon becomes very important, when the weak energy condition is violated. Such a violation is inevitable when quantum effects connected with the Hawking radiation are taken into account.  For example, there may exist a closed apparent horizon, while there is no event horizon \cite{Frolov:1981mz}. In the rest of the paper we discuss  possible "paradoxical" behavior of the event horizon in the case when the energy conditions are violated.

\section{Wormhole and entropy of Rindler horizon}\label{ra_sec1}

If there exist matter violating the energy conditions, the Einstein equations allow solutions which describe traversable wormholes  (a detailed discussion of this subject can be found in a nice book \cite{visser1996lorentzian}). Such wormholes can be used as a device for "study" of black hole interiors \cite{Frolov:1993jq} . When the mass and the size of mouths of the wormhole are small and the mass of the (non-rotating) black hole is large this effect allows a simple analytical description. We shall use an approach proposed by Emparan and collaborators for study of the event horizon for a small mass black hole falling into the very large black hole \cite{Emparan:2016ylg,Emparan:2016ipc,Emparan:2017vyp} . Namely, we approximate the event horizon of the black hole of large mass $M$ by the Rindler horizon. This is a plane null surface in a flat spacetime. In the presence of a static wormhole, this plane first passes one of the mouths, and some time later it passes the other mouth. We are going to describe what happens in such a process with the event horizon. Here and later we consider a case when the number of spacetime dimensions is 4 or higher. This does not make the problem more complicated, but it is useful for a general discussion.

Let us consider $(D=n+2)$-dimensional flat spacetime and denote by
\be\n{a.1}
X^a=(T,X,\vec{Z})\hh \vec{Z}=Z_1,\ldots ,Z_n\, ,
\ee
Cartesian coordinates in it. Denote by $\gamma_{\pm}$ two straight  lines  defined by the equations
\be\n{a.2}
\gamma_{\pm}: X=\pm {1\over 2} L\hh Z_i=0\, .
\ee
We assume that these two lines represent positions of the two spherical mouths of a traversable wormhole. We assume that the radii of the mouths are equal and very small, and that the time at the mouthes obeys the relation $T_+=T_-$.  In such a model, any null ray that enters into $\gamma_{-}$-mouth at time $T$ comes out  $\gamma_{+}$-mouth at the same moment of time.

We identify the unmodified Rindler horizon with a null plane $\Pi$: $T=X-L/2$. It crosses $\gamma_{+}$ at $T=0$.
We also denote by $N$ the past null cone with the vertex at $\gamma_{-}$-mouth at the moment of time $T=0$ (see Figure~1).

\begin{figure}
\centerline{\includegraphics[width=4.0in]{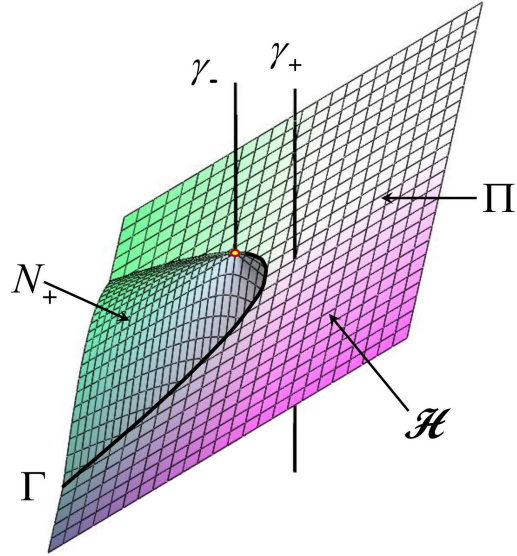}}
\caption{This picture illustrates how the surface of Rindler horizon is modified in the presence of a static traversable wormhole. At this picture time changes in the vertical direction. Two solid lines $\gamma_{\pm}$ represent positions of the mouths  of the traversable wormhole.} \label{ra_fig2}
\end{figure}

In our problem there exists only one dimensional parameter $L$ which gives the length scale. It is convenient to use dimensionless coordinates $x^a=X^a/L=(t,x,\vec{z})$. We also denote $\xi=x+1/2$. In these coordinates $(t,\xi,\vec{z})$ the equations of the past null cone $N$ and the undeformed horizon $\Pi$ are
\be
\xi^2+\rho^2=t^2\hh t=\xi-1\, .\n{a.3}
\ee
where $\rho=\sqrt{\vec{z}^2}.$
This null cone intersects the horizon $\Pi$ at $n$-dimensional surface $\Gamma$, which is a simultaneous solution of the equations (\ref{a.3})
\be
x=-{1\over 2}\rho^2\hh t=-{1\over 2}(\rho^2+1)\, .
\ee
We denote by $N_+$ a part of the cone $N$ locates above the null plane $\Pi$, and by ${\cal H}$ a part of the plane $\Pi$ located outside the surface $\Gamma$. It is easy to check that Rindler horizon, modified by the presence of the wormhole, is $H=N_+\cup {\cal H}$.
Really, an additional region lying between $N_+$ and $\Pi$ is visible in the presence of the wormhole. The rays from this domain can pass through the wormhole and appear outside  $\gamma_+$ mouth before the null plane $\Pi$ crosses it.

\begin{figure}[ht]
\centerline{
\minifigure[$\tau=1/2$]
{\includegraphics[width=2in]{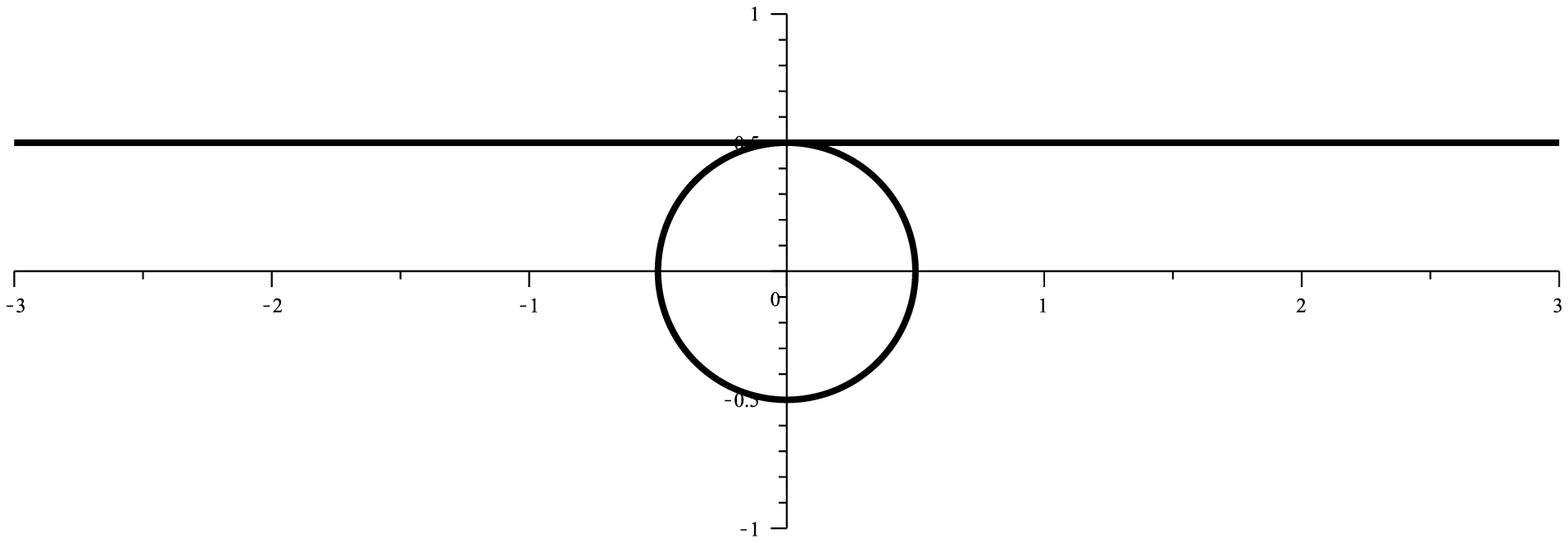}\label{AA_2}}
\hspace*{4pt}
\minifigure[$1/2<\tau<1$]
{\includegraphics[width=2in]{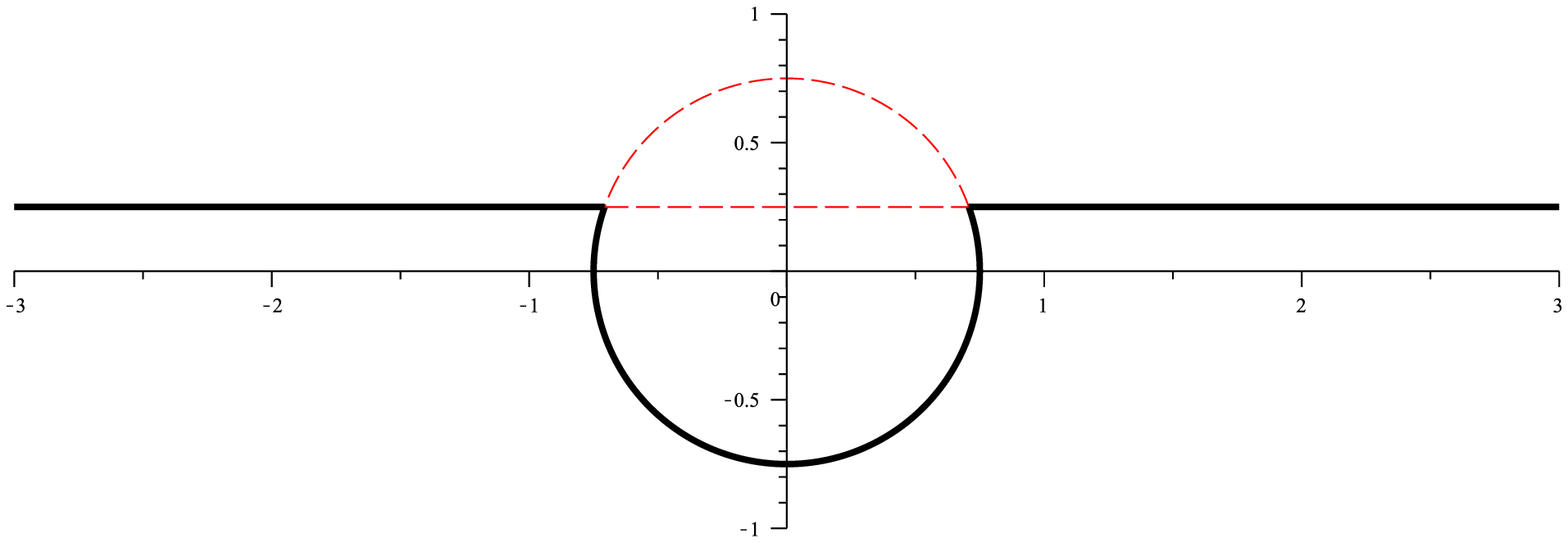}\label{AA_3}}
 }
\centerline{
\minifigure[$\tau=1$]
{\includegraphics[width=2in]{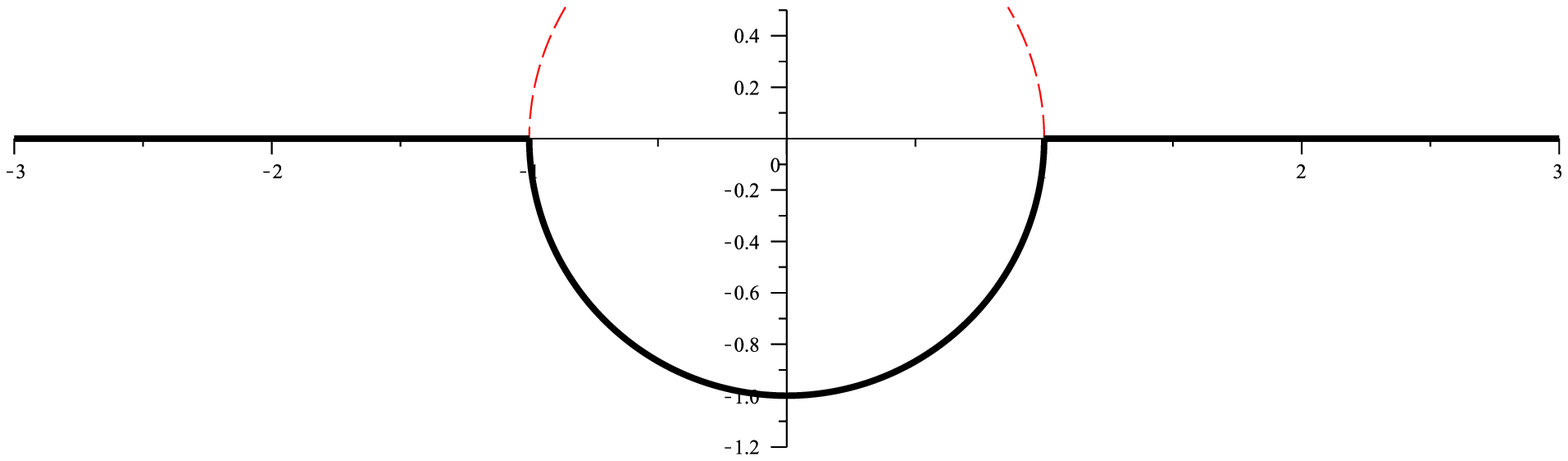}\label{AA_4}}
\hspace*{4pt}
\minifigure[$\tau >1$]
{\includegraphics[width=2in]{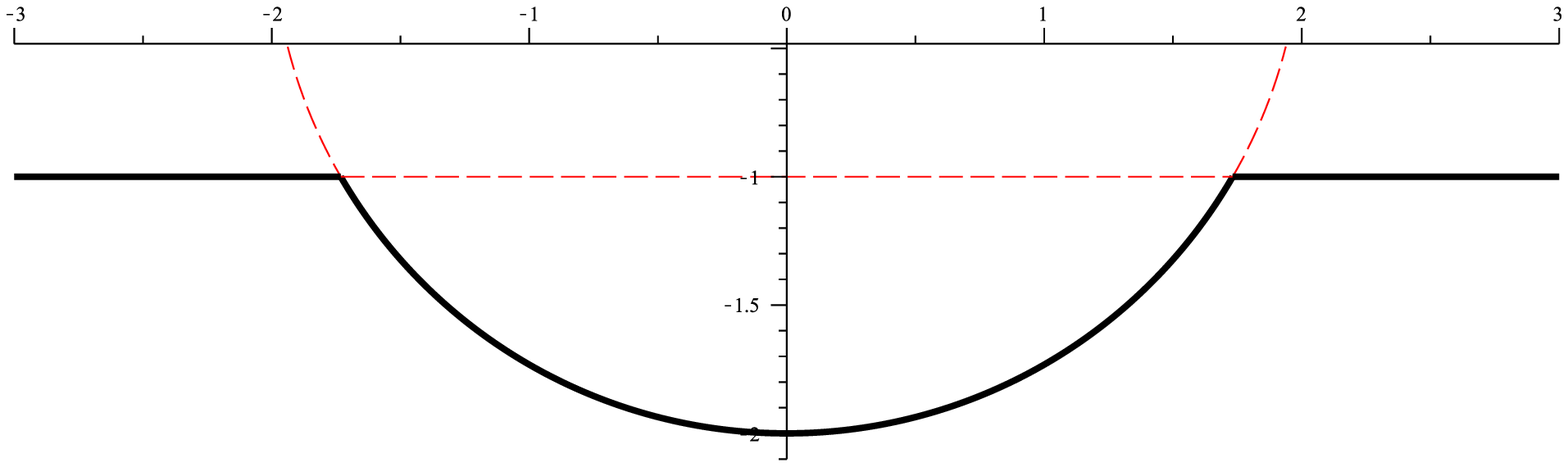}\label{AA_5}}
 }
 \caption{$t=-\tau=$const slices of the horizon}\label{AAA}
\end{figure}

Since interesting modification of the event horizon occurs at negative value of time $t$, it is convenient to define $\tau=-t$.
We denote by ${\cal B}_{\tau}$ a slice of the distorted Rindler horizon at time $t=-\tau$.
For $t\ge 0$  ${\cal B}_{\tau}$ is $n$-dimensional Euclidean spaces $E^n$. For $1/2 < \tau <1$, ${\cal B}_{\tau}$ slice is a sum of 2 disconnected $n$ dimensional spaces. One of them is $E^n$ and the other is a round sphere $S^n$ of radius $\tau$. The surface area of a unit $n$-dimensional sphere $S^n$ is
\be
S_n={2\pi^{n/2}\over \Gamma(n/2)}\, .
\ee
In a flat space  a sphere of radius $\tau$ contains a volume $V_{n+1}={1\over n}\tau^{n+1} S_n$\, .

In order to calculate the surface area of the horizon slices let us make some remarks. Let us denote by $d\sigma_n^2$ the line element of the surface of $n$ dimensional unit sphere
\be \n{sn2}
d\sigma_n^2=d\theta^2 +\sin^2\theta d\sigma_{n-1}^2 \, .
\ee
To illustrate the properties of the horizon slices at different moments of time we consider a section $\theta_1=\ldots \phi=0$, and chose the parameter $\theta$ so that its value $\theta=0$ corresponds to the lowest point of the sphere.

Let us denote
\be\n{ggg}
\cos \theta_{\Gamma}=1-{1\over \tau}\hh \mbox{  for   } \tau >1/2\, .
\ee
The angle $\theta_{\Gamma}$ determines a point where $\Gamma$ curve intersects the slice $t=-\tau=$const. In the interval $\tau\in [1/2,\infty)$, $\theta_{\Gamma}$ monotonically decreases from  $\pi$ at $\tau=1/2$ to $0$ when $\tau\to\infty$. For large $\tau$, $\theta_{\Gamma}\sim (2/\tau)^{1/2}$. The structure of slices ${\cal B}_{\tau}$ for $\tau\ge 1/2$ is shown at Figure~\ref{AAA}. For $\tau=1$, ${\cal B}_{1}$ is a half of the sphere $S^n$ of radius 1 glued to a part of $E^n$ (see Fig~\ref{AA_4}). For $\tau>1$ the slice has the form presented at Fig~\ref{AA_5}.

The line element of the spherical component of the event horizon at $\tau>0$ is
\be
dl^2=\tau^2 d\sigma_n^2\, .
\ee
The area of the distorted part of the event horizon is
\be
A_n(\tau)=\tau^n S_{n-1}P_n\hh P_n(\theta_{\Gamma})=\int_{0}^{\theta_{\Gamma}} d\theta \sin^{n-1}(\theta)\, .
\ee
The area of the part of the undistorted horizon located inside $\Gamma$ at time $\tau$ is
\be
A_n^0(\tau)={1\over n}\tau^n \sin^n(\theta_{\Gamma})S_{n-1}\, .
\ee
Thus, the change of the area of the horizon at $\tau>1/2$ is
\be
\Delta A_n=A_n^0(\tau)-A_n(\tau)= S_{n-1}\Delta\hh
\Delta={1\over n} (2\tau-1)^{n/2}-\tau^n P_n(\theta_{\Gamma})\, .
\ee

For the case $D=4$ ($n=2$) and $\tau\ge 1/2$, the area deficit $\Delta A_2$ does not depend on time  and it is
\be
\Delta A_2=-\pi\, .
\ee
For $\tau\in [0,1/2]$
\be
\Delta A_2=-4\pi \tau^2\, ,
\ee
so that at $\tau=0$, that is when the horizon is not distorted, $\Delta A_2$ vanishes.

The evolution of the entropy of the Rindler horizon in the presence of the wormhole can be described as follows. The Rindler event horizon is defined by the following condition: At late time in future it coincides with the plane $\Pi$. The surface of this horizon in the past is slightly different from this plane. This difference asymptotically vanishes in the infinite past, however the size of domain where the horizon differs from $\Pi$, grows in the past. As a combination of these two effects, $\Delta A_2$ remains the same. If we consider evolution of this "real" Rindler horizon, we come to the conclusion that it does not change up to the moment of time $t=-1/2$. At this moment a spherical part of the event horizon separates. Its initial radius is $1/2$ and its area is $\pi$. The area of the unperturbed horizon $A_n^0$ becomes less then the area of the initial horizon by the value $\pi$. Later in time the separated part of the horizon shrinks to zero. The net result of this evolution is that the area of the Rindler horizon at late time is less that its initial value by quantity $\pi$.
If we restore the dimensionality one gets the following expression for the decrease of the entropy
\be
\Delta S=S(t=\infty)-S(t=-\infty)=-\pi {L^2\over 4l_{pl}^2}\, .
\ee

If the number of spacetime dimensions is greater than four, the behavior of the entropy associated with the Rindler horizon is even more peculiar. For example for $D=5$ ($n=3$) one has
\ba
&&\Delta A_3=4\pi \Delta_3\, ,\\
&&\Delta_3={1\over 6} (\tau+1)(3\tau-2)\sqrt{2\tau-1}-{1\over 2}\tau^3\arccos({\tau-1\over \tau})\, .
\ea
For $\tau\to \infty$ $\Delta A_3$ is negative and its absolute value infinitely grows
\be
\Delta A_3\sim \left[-{2\sqrt{2}\over 5} \tau^{1/2}+{\sqrt{2}\over 14}\tau^{-1/2}\right] S_2\, .
\ee

Calculations for $D=6$ give
\be
\Delta A_4=2\pi^2 \Delta_4\hh \Delta_4={1\over 4}-{2\over 3}\tau\, .
\ee
Similar results are valid for any $n\ge 3$: the corresponding $\Delta A_n$ is negative and divergent in the limit $\tau\to \infty$.

\section{A wormhole in a black hole geometry}

As we mentioned, the Rindler horizon is a certain idealization of the black hole horizon in the limit when the mass of the latter is very large. However, there exit two properties that differ the Rindler horizon from the horizon of the black hole: its topology differs from the spherical topology of the black hole, and its surface area is infinite. The first difference implies that the Rindler horizon is not a trapped surface. In this section we discuss how the black-hole event horizon evolves in the presence of a traversable wormhole, when one of its mouths falls into it.

\subsection{Null rays in the black-hole interior}

We write the metric of $(D=n+2)$-dimensional spherical black hole in the form
\be
dS^2=-f dV^2+2 dV dr+r^2 d\omega_n^2\, ,
\ee
where $f=1-(r_g/r)^{n-1}$. In is convenient to use the dimensionless form of the coordinates and the metric. For this purpose we put
\be
V=r_g v\hh r= r_g x\hh dS^2=r_g^2 ds^2\, ,
\ee
where
\be
ds^2=-f dv^2 +2 dv dx+x^2 d\omega_n^2\hh f=-{1-x^{n-1}\over x^{n-1}}\, .
\ee
In the black hole interior $x\in (0,1)$ the function $f$ is negative.

We consider a situation when one of the mouths of a traversable wormhole falls down into the black hole, while the other mouth is kept at rest in its exterior. As earlier, we assume that the size of the mouths is very small. We also assume that the inner mouth reaches the singularity at the moment $v=0$ of the advanced time. All the causal curves that enter the inner mouth before this would appear in the black hole exterior. We denote by $N$ the past null cone with the vertex at $x=0$, that is in the position of the inner mouth at  $v=0$. We also denote by $\Gamma$ an intersection of $N$ with the surface $x=1$, which is a position of unperturbed horizon.
We denote by $N_+$ a part of $N$ with $x\le 1$ restricted by $\Gamma$, and by ${\cal H}$ a part of the horizon $H$ outside $\Gamma$, that is an undistorted part of the horizon. Thus the surface of the distorted horizon is $N_+\cup {\cal H}$.

To find the $n$-dimensional surface $\Gamma$ one needs to solve equations for null rays in the black hole geometry. Let us write the line element of a unit sphere in the form
\be
d\omega_n^2=d\theta^2 +\sin^2\theta d\omega_{n-1}^2\, .
\ee
Because of the spherical symmetry it is sufficient to consider only null rays, propagating along the axis of the symmetry of $S^{n-1}$.
In the coordinates $(v,x,\theta,\theta_i)$, with $i=1,\ldots ,n-2$ one can put $\theta_i=0$. Thus a null ray from the cone $N$ can be specified by functions $(v(\nu), x(\nu), \theta(\nu))$, where $\nu$ is an affine parameter along the ray.  The problem reduces to finding null rays in 3D spacetime with metric
\be
ds^2=-f dv^2+2 dv dx +x^2 d\theta^2\, .
\ee
It has 2 Killing vectors $\partial_v$ and $\partial_\theta $, hence one has
\ba
&& -f \dot{v}+\dot{x}=E\, ,\n{eq1}\\
&&x^2 \dot{\theta}=L\, ,\n{eq2}\\
&&-f \dot{v}^2+2 \dot{v}\dot{x}+x^2\dot{\theta}^2=0\, .\n{eq3}
\ea
The first two relations correspond to the conservation of the energy $E$ and the angular momentum $L$. The last relation is a normalization condition for the null ray velocity. One can exclude the parameter $E$. It is sufficient to change the affine parameter $\nu\to \tilde{\nu}=E^{-1}\nu$.
We use a dot to denote  a derivative with respect to this new parameter.

To determine a position of the distorted horizon one needs to trace back in time the beam of null rays that arrive to the inner mouth of the wormhole at the moment when it  reaches the singularity, that is a point $(v=0,x=0,\theta=0)$. The null rays from such a beam differs by the value of the parameter $\lambda=L/E$. Inside the black hole one has $f=-|f|$ and the
equations (\ref{eq1})--(\ref{eq3}) take the form
\ba
&&\dot{x}=1-|f| \dot{v}\, ,\n{xxx}\\
&&\dot{\theta}={\lambda\over x^2}\, ,\n{th}\\
&& -|f|\dot{v}^2+2\dot{v} +{\lambda^2\over x^2}=0\, .
\ea
The last equation gives
\be\n{vdot}
\dot{v}={1-\epsilon U \over |f|}\hh U=\sqrt{1+\lambda^2 |f|/x^2}\, .
\ee
The parameter $\epsilon$ takes the values $\pm 1$. For $\lambda=0$  the equation (\ref{vdot}) takes the form
\be
\dot{v}={1-\epsilon\over |f|}\, .
\ee
Hence for $\epsilon=+1$, $\dot{v}=0$. Such ray  propagates along fixed advanced time surface. For the other ray with $\lambda=0$, one has $\dot{v}=2/|f|$. Such a ray propagates along fixed "retarded time" surface. Only the rays  with $\epsilon=+1$, after being traced back in time, cross the event horizon.

\subsection{$\Gamma -$curve}

A curve $\Gamma$ is formed by intersection of the rays with $\epsilon=1$  with the undistorted horizon. To find $\Gamma$ one needs to solve the following equations
\be
\dot{v}={1-U\over |f|}\hhh \dot{x}=U\hhh \dot{\theta}={\lambda\over x^2}\, .
\ee
These equations imply
\be\n{vtx}
{dv\over dx}={ 1-U\over |f| U}\equiv -{\lambda^2\over x^2 U (1+U)}\hh {d\theta\over dx}={\lambda\over x^2 U}\, .
\ee
The function $U$ is singular at $x=0$. The right-hand side of these equations can be written in a more clear form. Let us denote
\be
U={{\cal U}\over x^{ {n+1\over 2}}}\hh {\cal U}=\sqrt{\lambda^2 (1-x^{n-1})+x^{n+1}}\, .
\ee
Then the equations (\ref{vtx}) takes the form
\be\n{vvttx}
{dv\over dx}= -{\lambda^2 x^{n-1}\over {\cal  U} ({\cal  U}+x^{ {n+1\over 2}})}\hh
{d\theta\over dx}={\lambda x^{ {n-3\over 2}}\over {\cal  U}}\, .
\ee

The quantity $\lambda$ enters as a parameter in the right-hand side of these equation. It is easy to find solutions of (\ref{vvttx}) for special limiting value of this parameter.
Namely for $\lambda=0$ one has $v(x)=\theta(x)=0$. For $\lambda\to \infty$,  ${\cal U}\sim \lambda \sqrt{1-x^{n-1}}$ and equations (\ref{vvttx}) take the form
\be\n{inf}
{dv\over dx}= -{x^{n-1}\over 1-x^{n-1}}\hh
{d\theta\over dx}={x^{ {n-3\over 2}}\over \sqrt{1-x^{n-1}}}\, .
\ee
The solution of the second equations is
\be
\theta={2\over n-1}\arcsin(x^{-{n-1\over 2}})\, .
\ee
In particular, the value  $\theta_{\infty}$ of the angle $\theta$ at the moment when a null ray with the impact parameter $\lambda=\infty$ reaches the unperturbed horizon $x=1$ is
\be
\theta^{\Gamma}_{\infty}={\pi\over n-1}\, .
\ee
The corresponding advanced time $v^{\Gamma}_{\infty}=-\infty$.

For a finite value of the impact parameter $\lambda$ the coordinates $(v_{\Gamma},\theta_{\Gamma})$ of the points, where the ray  crosses the unperturbed horizon $x=1$, are
\ba
&& v_{\Gamma}(\lambda)=-\lambda^2 \int_0^1 dx {x^{n-1}\over {\cal U} ({\cal U}+x^{ {n+1\over 2}})}\, ,\\
&& \theta_{\Gamma} (\lambda)= \lambda \int_0^1 dx  {   x^{{n-3\over 2}} \over  {\cal  U} }\, .
\ea
These relations are nothing but parametric equations of the surface $\Gamma$.

\subsection{Shape and area of the distorted event horizon}

In order to determine a shape of the event horizon at some moment of time one needs to integrate backward in time $v$ null ray equation for both subsets of the rays, with $\epsilon=\pm 1$. We need to find intersection of such rays with $v_0=$const surface, with $v_0<0$. For the rays from the subset $\epsilon=-1$ the corresponding range of $\lambda$ is $\lambda\in [0,\infty)$. For the other subset of rays with $\epsilon=+1$,  $\lambda$ is in the interval $\lambda\in [\lambda_{\Gamma}(v_0),\infty)$. $\lambda_{\Gamma}(v_0)$ is the impact parameter of the ray that crosses $\Gamma$  at $v=v_0$. To make the calculations easier we introduce a new parameter $\mu=\epsilon/ \lambda$, and denote $\mu_{\Gamma}=1/\lambda_{\Gamma}$. Then $\mu$ changes from $-\infty$ till $\mu_{\Gamma}$. The negative $\mu$ corresponds to a subset with $\epsilon=-1$, while the positive $\mu$ correspond to the other subset. We also denote $y=x^n$.
Then the corresponding set of null-ray equations takes the form
\ba
&& {dy\over dv}=-n V W\hh {d\theta\over dv}=-{W\over y^{ {n+1\over 2n}}}\, ,\\
&& V=\sqrt{1-y^{{n-1\over n}}+\mu^2 y^{{n+1\over n}}}\hh
W=V+\mu y^{{n+1\over 2n}}\, .\n{eqmot}
\ea
The initial conditions  at $v=0$ are $y=\theta=0$.

At $y=0$ one has $V(0)=W(0)=1$. The first, $y-$equation is regular at this point, while the second equation has integrable singularity. In order to treat this singularity it is sufficient to write $\theta=\tilde{\theta}+\Delta\theta(y)$ and to chose the (finite at $y=0$) function $\Delta\theta$ so that the equation for $\tilde{\theta}$ becomes regular at $y=0$. One has
\be
\Delta\theta=\left\{
\begin{array}{cl}
2y^{1/4}+{1\over 3}  y^{3/4}\, ,  & \mbox{if  $n=2$ }\, , \\
{2\over n-1} y^{ {n-1\over 2n}}\, ,   &  \mbox{if  $ n\ge 3$}\, .
\end{array}
\right.
\ee
The equation for $\tilde{\theta}$ is
\be\n{tT}
 {d\tilde{\theta}\over dv}=\tilde{\Theta}
 \hh
 \tilde{\Theta}=\Theta-{d\Delta\theta\over dy} Y\, .
\ee
For $n\ge 3$
\be
\tilde{\Theta}= -{W\over y^{ {n+1\over 2n}}}(1-V)\, .
\ee
It is easy to check that for $\tilde{\Theta}\sim y^{ {n-3\over 2 n}}$, so that that the right-hand side of (\ref{tT}) is finite at $y=0$.

\begin{figure}[ht]
\centerline{
\minifigure[$v_0=-0.05$]
{\includegraphics[width=1.8in]{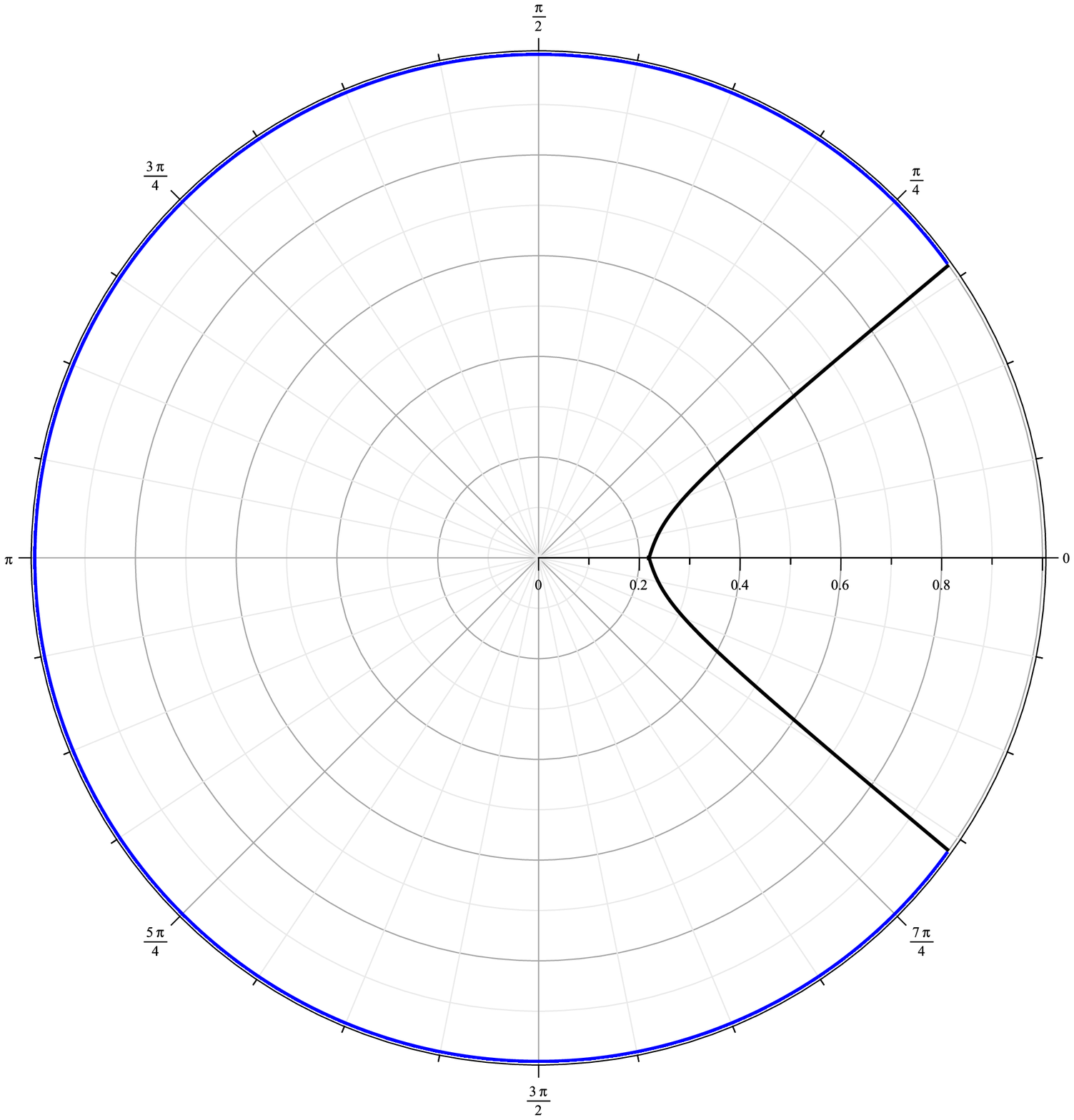}}
\hspace*{4pt}
\minifigure[$v_0=-0.1$]
{\includegraphics[width=1.8in]{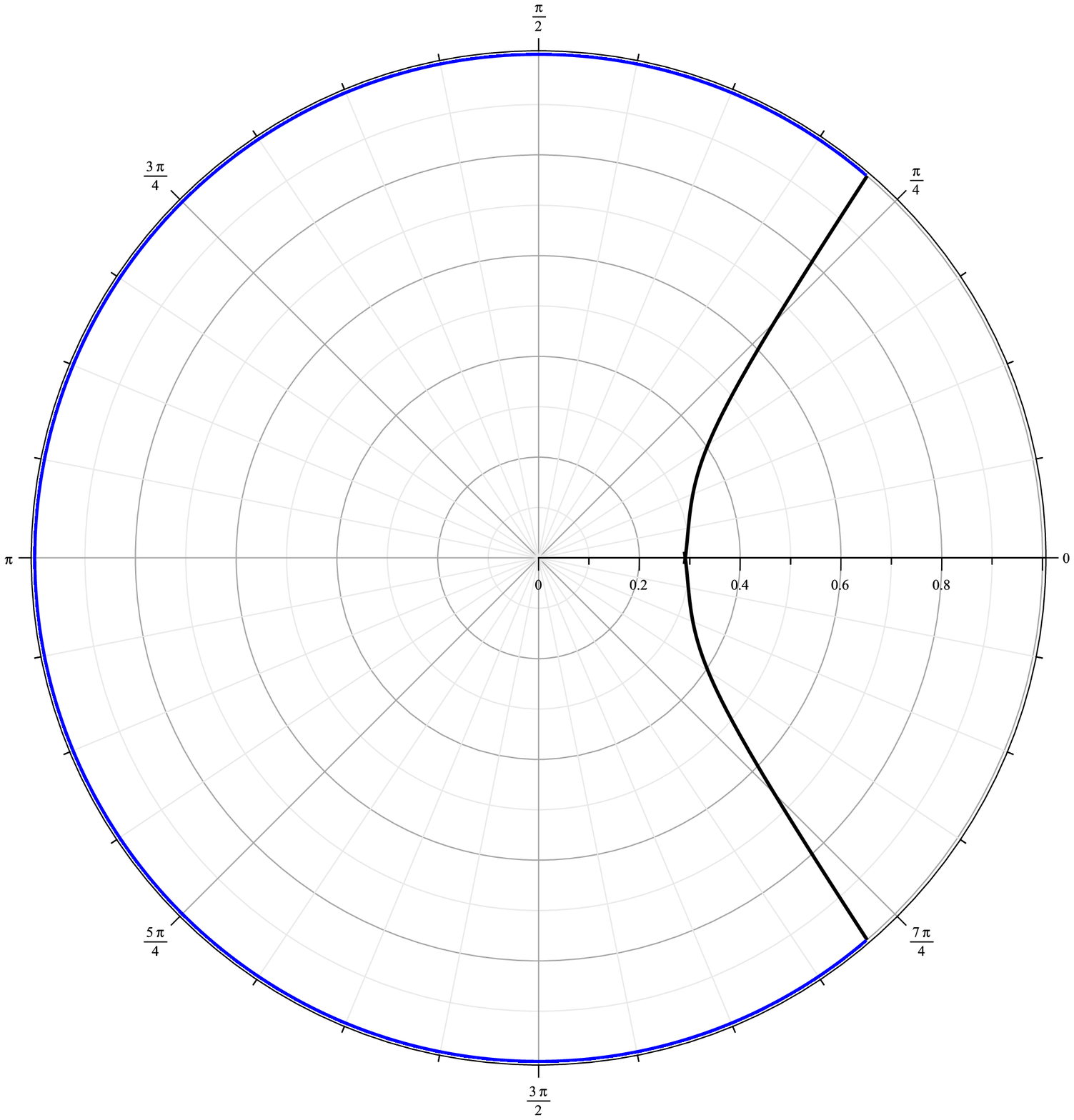}}
 }
\centerline{
\minifigure[$v_0=-0.5$]
{\includegraphics[width=1.8in]{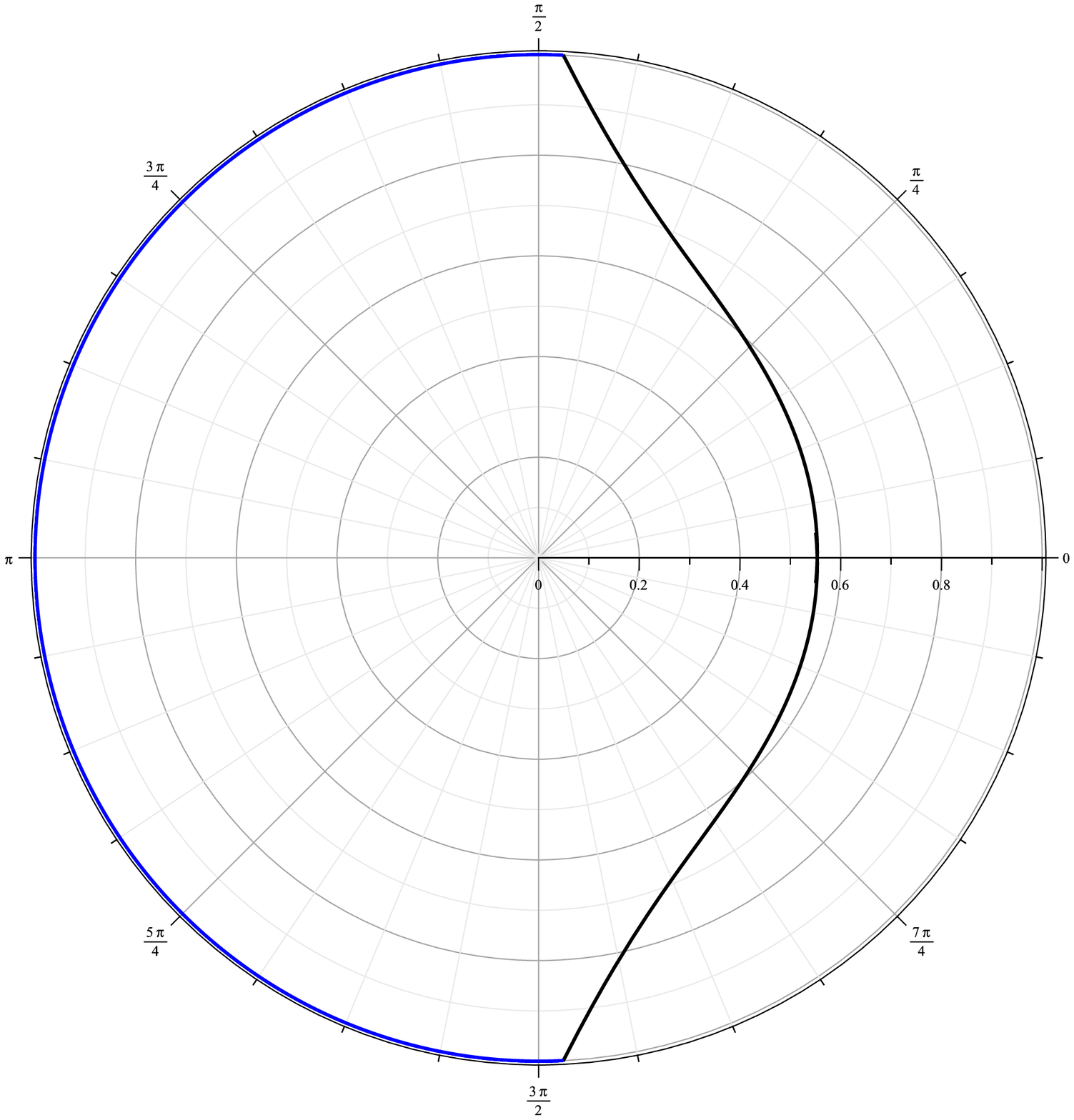}}
\hspace*{4pt}
\minifigure[$v_0=-1.0$]
{\includegraphics[width=1.8in]{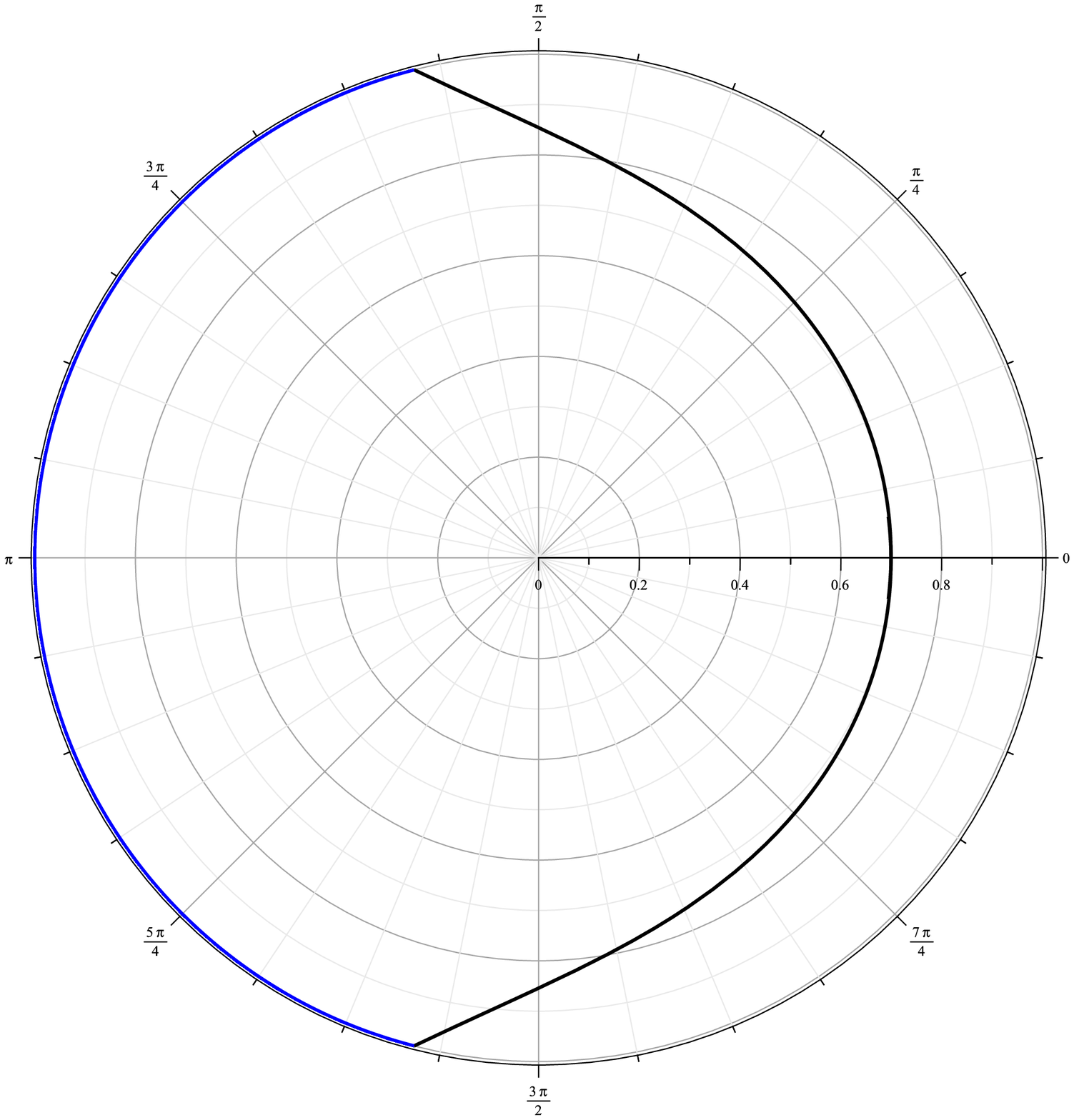}}
 }
 \caption{Shape of the horizon slices at different moments of the advanced time $v_0$.}\label{PPP}
\end{figure}

By integrating the equations (\ref{eqmot}) till the moment $v_0$ of the advanced time $v$, one obtains values $x_0(\mu)=x(v_0,\mu)$ and $\theta_0(\mu)=\theta_0(v_0,\mu)$. These expressions determine parametrically the shape of the $v=v_0$ slice of the deformed  horizon. The full slice of the deformed horizon corresponds to the values of $\mu$ in the interval $\mu\in (-\infty,\mu_{\Gamma}]$. The point $\mu_{\Gamma}$, where $x_0=x_0(\mu_{\Gamma})=1$ and $\theta_0=\theta_0(\mu_{\Gamma})$, is a point of $\Gamma$ where the slice $v_0$  intersects the unperturbed horizon.

The Figures~\ref{PPP} shows the coordinate form of the slices $v=v_0$ of the event horizon for different values of the advanced time $v_0$.
In these diagrams we used polar coordinates $(x,\theta)$. By rotation of the plots around the horizontal line one obtains a two-dimensional surface illustrating the form of the slice $v=$const of the event horizon. It should be emphasized that the two-geometry induced on such a rotation surface by its embedding into 3-dimensional flat space does not coincide with the two geometry of a slice of the distorted horizon. The latter is
\be
d\omega^2=x^2(\theta) (d\theta^2+\sin^2\theta d\phi^2)\, .
\ee
Namely this metric determines the area element of the horizon slice, $d\sigma= x^2(\theta) \sin(\theta) d\theta d\phi$.
The surface area of the deformed slice of the horizon is
\be \n{AR}
A_n(v_0)=r_g^n S_{n-1} P_n[x]\hh
P_n[x]=\int_0^{\theta_{\Gamma}} d\theta \sin^{n-1}(\theta) x^n(\theta)\, .
\ee
The same expression with $x(\theta)=1$ give the value of the surface area $A_n^0(v_0)$ of the unperturbed part $0\le\theta\le \theta_{\Gamma}$ of the horizon. The difference
\be
\Delta A_n(v_0)=A_n^0(v_0)-A_n(v_0)
\ee
determined the total difference of the areas of the undistorted and distorted horizons.
The total area of undistorted horizon is $A_H^n=r_g^n S_n$.

The relative change of the horizon area  $\Delta_n=\Delta A_n(v_0)/A_H^n$ is
\be
\Delta_n={P_n[1]-P_n[x]\over B_n} \hh B_n=\int_0^{\pi} d\theta \sin^{n-1}(\theta)\, .
\ee

Omitting technical details, we just present here the results of the numerical calculations of the quantity $\Delta_n$ for the physically most interesting case $n=4$ ($D=4$). For a small absolute value of negative $v_0$  $\Delta_2$ is an increasing function of $|v_0|$. Its values are: $0.2$ for $v_0=-0.01$; $0.35$ for $v_0=-0.05$; and $0.41$ for $v_0=-0.1$. Near $v_0\approx -0.2$ $\Delta_2$ reaches the maximum $\approx 0.43$ and it further decreases again: $0.36$ for $v_0=-0.5$; $0.28$ for $v_0=-1.0$; and $0.1$ for $v_0=-2.0$.

If we relate the black hole entropy to its surface area according to (\ref{BHBH}), we would arrive to the following conclusion. In the above described gedanken experiment with a wormhole, the black hole entropy depends on the advanced time $v$. At first it decreases. At some moment $v<0$ it reaches the minimal value and after this increases again. The black hole entropy returns back to its original value at the moment of advanced time $v=0$, when the inner mouth of the wormhole is crashed in the singularity. The maximal decrease of the entropy reaches  $43\%$ of its original value. Similar results are valid for higher dimensional black hole.

We found the above described situation rather paradoxical. Even in the presence of a tiny mass wormhole, which practically does not change the mass and gravitational field of the black hole, there exists a process in which the black hole entropy decreases by $43\%$ of its original value. The time scale of this process is $\sim r_g/c$.  It is natural to relate this change of the entropy to the fact that during the time when the inner mouth is falling in the black hole interior, the information about the black hole interior becomes available to the external observer. At any rate, this dramatic decrease of the entropy is not expected in the "normal" thermodynamical system, without a dramatic change of their parameters.
There might exist an interesting possible resolution of this "paradox". In 1990 Jacob Bekenstein discussed quantum limitations on the storage and transmission of  information \cite{Bekenstein:1990du} . It might happen that the above assumption that by using of the wormhole with mouths of arbitrary small size one can extract all the information from a spacetime domain, which geometrically became available, is wrong. If the rate of the information extraction is limited by the geometric characteristics of the wormhole, this idealization would be oversimplified. An interesting question is how to measure the black hole entropy in a situation when the area of the black hole changes fast, and what are requirements on the devices used in such measurements.

\section*{Acknowledgements}
\label{sc:acknowledgements}
\addcontentsline{toc}{section}{Acknowledgements}

The author thanks the Natural Sciences and Engineering Research Council of Canada (NSERC) and the Killam Trust for their financial support.

\bibliographystyle{ws-rv-van}


\end{document}